# The Future Has Thicker Tails than the Past: Model Error As Branching Counterfactuals

**Nassim N Taleb**
NYU-Poly Institute

## Abstract

*Ex ante* forecast outcomes should be interpreted as counterfactuals (potential histories), with errors as the spread between outcomes. Reapplying measurements of uncertainty about the estimation errors of the estimation errors of an estimation leads to branching counterfactuals. Such recursions of epistemic uncertainty have markedly different distributial properties from conventional sampling error. Nested counterfactuals of error rates invariably lead to fat tails, regardless of the probability distribution used, and to powerlaws under some conditions. A mere .01% branching error rate about the STD (itself an error rate), and .01% branching error rate about that error rate, etc. (recursing all the way) results in explosive (and infinite) higher moments than 1. Missing any degree of regress leads to the underestimation of small probabilities and concave payoffs (a standard example of which is Fukushima). The paper states the conditions under which higher order rates of uncertainty (expressed in spreads of counterfactuals) alters the shapes the of final distribution and shows which *a priori* beliefs about conterfactuals are needed to accept the reliability of conventional probabilistic methods (thin tails or mildly fat tails).

## Introduction

**Intuition.** An event has never shown in past samples; we are told that it was "estimated" as having zero probability. But an estimation has to have an error rate; only measures deemed *a priori* or fallen from the sky and dictated by some infaillible deity can escape such error. Since probabilities cannot be negative, the estimation error will necessarily put a lower bound on it and make the probability > 0.

This, in a nutschell, is how we should treat the convexity bias stemming from uncertainty about small probabilities. Using the same reasoning, we need to increase the raw "estimation" of small probabilities. There can be uncertainty about the relationship between past samples and future ones, or, more philosophically, from the problem of induction. Doubting the reliability of the methods used to produce these probabilities, or doubting beliefs about the future resembling the past will lend us to exercise standard skepticism and consider a spate of different alternative future outcomes. The small probability event will necessarily have, in expectation, i.e., on average across all potential future histories, a higher than what was measured. The increase in the probability will be commensurate with the error rate in the estimation. It, simply, results from the convexity bias that makes small probabilities rise when we are uncertain about them. Accordingly, the future needs to be dealt with as having thicker tails than what was measured in the past.

**Incoherence in Probabilistic Measurements.** Just as "estimating" an event to be of measure 0 is incoherent, it is equally inconsistent to estimate anything without introducing an estimation error in the analysis and adding a convexity bias (positive or negative). But this incoherence (or confusion between estimated and *a priori*) pervades the economics literature whenever probabilistic and statistical methods are used. For instance, the highest use of probabiltiy in modern financial economics resulting from the seminal Markowitz (1952) has his derivations starting with assuming E and V (expectation and variance). At the end of the paper he states that these parameters need to be estimated. Injecting an estimation error in the analysis entirely cancel the derivations of the paper as they are based on immutable certainties.

## Regressing Counterfactuals

We can go beyond probabilities and, in place of probabilities, perturbate parameters of probability distributions used in practice and perturbate the rates of perturbation. So this paper introduces two notions: treating errors as branching counterfactual histories and regressing (i.e., compounding) the error rates. An error rate about a forecast can be estimated (or, of course, "guessed"). The estimation (or "guess"), in turn, will have an error rate. The estimation of such error rate will have an error rate. (The forecast can be an economic variable, the future rainfall in Brazil, or the damage from a nuclear accident).

What is called a regress argument by philosophers can be used to put some scrutiny on quantitative methods or risk and probability. The mere existence of such regress argument will lead to series of branching counterfactuals three different regimes, two of which lead to the necessity to raise the values of small probabilities, and one of them to the necessity to use power law distributions. This study of the structures of the error rates refines the analysis of the *Fourth Quadrant* (Taleb, 2008) setting the limit of the possibility of the use of probabilistic methods (and their reliability in the decision-making), based on errors in the tails of the distribution.

So the boundary between the regimes is what this paper is about -what assumptions one needs to have set beforehand to avoid radical skepticism and which specific *a priori* undefeasable beliefs are necessary to hold for that. In other words someone using probabilistic estimates should tell us beforehand which immutable certainties are built into his representation, and what should be subjected to error -and regress --otherwise they risk falling into a certain form of incoherence: if a parameter is estimated, second order effects need to be considered.

This paper can also help setting a wedge between forecasting and statistical estimation.

## The Regress Argument (Error about Error)

The main problem behind *The Black Swan* is the limited understanding of model (or representation) error, and, for those who get it, a lack of understanding of second order errors (about the methods used to compute the errors) and by a regress argument, an inability to continuously reapplying the thinking all the way to its limit (**particularly when they provide no reason to stop**). Again, I have no problem with stopping the recursion, provided it is accepted as a declared *a priori* that escapes quantitative and statistical methods. Also, few get the point that the skepticism in *The Back Swan* it does not invalidate all measurements of probability; its value lies in showing a map of domains that are vulnerable to such opacity, defining these domains based on their propensity to fat-tailedness (of end outcomes), sensitivity to convexity effects, and building robustness (i.e., mitigation of tail effects) by appropriate decision-making rules.

*Philosophers and Regress Arguments*: I was having a conversation with the philosopher Paul Boghossian about errors in the assumptions of a model (or its structure) not being customarily included back into the model itself when I realized that only a philosopher can understand a problem to which the entire quantitative field seems blind. For instance, probability professionals do not include in the probabilistic measurement itself an error rate about, say, the estimation of a parameter provided by an expert, or other uncertainties attending the computations. This would only be acceptable if they consciously accepted such limit.

Unlike philosophers, quantitative risk professionals ("quants") don't seem to remotely get regress arguments; questioning all the way (without making a stopping assumption) is foreign to them (what I've called scientific autism, the kind of scientific autism that got us into so many mistakes in finance and risk management situations such as the problem with the Fukushima reactor). Just reapplying layers of uncertainties may show convexity biases, and, fortunately, it does not necessarily kill probability theory; it just disciplines the use of some distributions, at the expense of others --distributions in the $\mathcal{L}^2$ norm (i.e., square integrable) are no longer valid, for epistemic reasons. This is does not mean we cannot have parametric distributions; it just means that when there is no structure to the error rates we need to stay in the power law domains, even if the data does not give us reasons for that.

The bad news is that this recursion of the error rate invalidates all common measures of small probabilitites --and has the effect of raising them.

Indeed, the conversation with the philosopher was quite a relief as I had a hard time discussing with quants and risk persons the point that *without understanding errors, a measure is nothing* and one should take the point to its logical consequence that *any measure of error needs to have its own error taken into account*.

*The epistemic and counterfactual aspect of standard deviations*: One sad conversation took place a decade ago with another academic, a professor of risk management who writes papers on Value at Risk (and still does): he could not get that the standard deviation of a distribution *for future outcomes* (and not the sampling of some properties of existing population), the measure of dispersion, needs to be interpreted as the measure of uncertainty, distance between counterfactuals, hence *epistemic*, and that it, in turn, should necessarily have uncertainties (errors) attached to it (unless he accepted infallibility of belief in such measure). One needs to look at the standard deviation -or other measures of dispersion -as a degree of ignorance about the future realizations of the process. The higher the uncertainty, the higher the measure of dispersion (variance, mean deviation, etc.)

Such uncertainty, by Jensen's inequality, creates non-negligible convexity biases. So far this is well known in places in which subordinated processes have been used --for instance stochastic variance models --but I have not seen the layering of uncertainties taken into account.

*Betting on Rare Events:* Finally, this note explains my main points about betting on rare events. Do I believe that the probability of the event is higher? No. I believe that any additional layer of uncertainty raises the payoff from such probability (because of convexity effects); in other words, a mistake in one direction is less harmful than the benefits in the other --and this paper has the derivations to show it.

## *Note: Counterfactuals, Estimation of the Future v/s Sampling Problem*

Note that it is hard to escape higher order uncertainties, even outside of the use of counterfactual: even when sampling from a conventional population, an error rate can come from the production of information (such as: is the information about the sample size correct? is the information correct and reliable?), etc. These higher order errors exist and could be severe in the event of convexity to parameters, but they are qualitatively different with forecasts concerning events that have not taken place yet.

This discussion is about an epistemic situation that is markedly different from a sampling problem as treated conventionally by the statistical community, particularly the Bayesian one. In the classical case of sampling by Gosset ("Student", 1908) from a normal distribution with an unknown variance (Fisher, 1925), the Student T Distribution (itself a power law) arises for the estimated mean since the square of the variations (deemed Gaussian) will be Chi-square distributed. The initial situation is one of completely unknown variance, but that is progressively discovered through sampling; and the degrees of freedom (from an increase in sample size) rapidly shrink the tails involved in the underlying distribution.

The case here is the exact opposite, as we have an a priori approach with no data: *we start with a known priorly estimated or "guessed" standard deviation, but with an unknown error on it expressed as a spread of branching outcomes*, and, given the a priori aspect of the exercise, we have no sample increase helping us to add to the information and shrink the tails. We just deal with nested counterfactuals.

Note that given that, unlike the Gosset's situation, we have a finite mean (since we don't hold it to be stochastic and know it a priori) hence we necessarily end in a situation of finite first moment (hence escape the Cauchy distribution), but, as we will see, a more complicated second moment.

See the discussion of the Gosset and Fisher approach in Chapter 1 of Mosteller and Tukey (1977). [I thank Andrew Gelman and Aaron Brown for the discussion].

# Main Results

Note that unless one stops the branching at an early stage, all the results raise small probabilities (in relation to their remoteness; the more remote the event, the worse the relative effect).

1. Under the regime of proportional constant (or increasing) recursive layers of uncertainty about rates of uncertainty, the distribution has infinite variance, even when one starts with a standard Gaussian.
2. Under the other regime, where the errors are decreasing (proportionally) for higher order errors, the ending distribution becomes fat-tailed but in a benign way as it retains its finite variance attribute (as well as all higher moments), allowing convergence to Gaussian under Central Limit.
3. We manage to set a boundary between these two regimes.
4. In both regimes the use of a thin-tailed distribution is not warranted unless higher order errors can be completely eliminated *a priori*.

*Epistemic not statistical re-derivation of power laws*: Note that previous derivations of power laws have been statistical (cumulative advantage, preferential attachment, winner-take-all effects, criticality), and the properties derived by Yule, Mandelbrot, Zipf, Simon, Bak, and others result from structural conditions or breaking the independence assumptions in the sums of random variables allowing for the application of the central limit theorem. This work is entirely epistemic, based on standard philosophical doubts and regress arguments.

# Methods and Derivations

### *Layering Uncertainties*

The idea is to hunt for convexity effects from the layering of higher order uncertainties (Taleb, 1997).

Take a standard probability distribution, say the Gaussian. The measure of dispersion, here $\sigma$, is estimated, and we need to attach some measure of dispersion around it. The uncertainty about the rate of uncertainty, so to speak, or higher order parameter, similar to what called the "volatility of volatility" in the lingo of option operators (see Taleb, 1997, Derman, 1994, Dupire, 1994, Hull and White, 1997) --here it would be "uncertainty rate about the uncertainty rate". And there is no reason to stop there: we can keep nesting these uncertainties into higher orders, with the uncertainty rate of the uncertainty rate of the uncertainty rate, and so forth. There is no reason to have certainty anywhere in the process.

Now, for that very reason, this paper shows that, in the absence of knowledge about the structure of higher orders of deviations, we are forced to use a power-law tails. Most derivations of power law tails have focused on processes (Zipf-Simon preferential attachment, cumulative advantage, entropy maximization under constraints, etc.) Here we just derive them using lack of knowledge about the rates of knowledge.

### *Higher order integrals in the Standard Gaussian Case*

We start with the case of a Gaussian and focus the uncertainty on the assumed standard deviation. Define $\phi(\mu,\sigma,x)$ as the Gaussian density function for value $x$ with mean $\mu$ and standard deviation $\sigma$.

A $2^{nd}$ order stochastic standard deviation is the integral of $\phi$ across values of $\sigma \in \,]0,\infty[$, under the measure $f(\overline{\sigma}, \sigma_1, \sigma)$, with $\sigma_1$ its scale parameter (our approach to trach the error of the error), not necessarily its standard deviation; the expected value of $\sigma_1$ is $\overline{\sigma_1}$.

$$f(x)_1 = \int_0^\infty \phi(\mu, \sigma, x) f(\overline{\sigma}, \sigma_1, \sigma) \, d\sigma \qquad (1)$$

Generalizing to the N$^{th}$ order, the density function *f(x)* becomes

$$f(x)_N = \int_0^\infty \ldots \int_0^\infty \phi(\mu, \sigma, x) f(\overline{\sigma}, \sigma_1, \sigma) \, f(\overline{\sigma_1}, \sigma_2, \sigma_1) \ldots f(\overline{\sigma_{N-1}}, \sigma_N, \sigma_{N-1}) \, d\sigma \, d\sigma_1 \, d\sigma_2 \ldots d\sigma_N \qquad (2)$$

The problem is that this approach is parameter-heavy and requires the specifications of the subordinated distributions (in finance, the lognormal has been traditionally used for $\sigma^2$ (or Gaussian for the ratio $\text{Log}[\frac{\sigma_t^2}{\sigma^2}]$ since the direct use of a Gaussian allows for negative values). We would need to specify a measure *f* for each layer of error rate. Instead this can be approximated by using the mean deviation for $\sigma$, as we will see next.

Note that branching variance does not always result in higher Kurtosis (4th moment) compared to the Gaussian; in the case of N-2, using the Gaussian and stochasticzing both $\mu$ and $\sigma$ will lead to bimodality the lowering of the 4th moment.

### *Discretization using nested series of two-states for $\sigma$- a simple multiplicative process*

A quite effective simplification to capture the convexity, the ratio of (or difference between) $\phi(\mu,\sigma,x)$ and $\int_0^\infty \phi(\mu, \sigma, x) f(\overline{\sigma}, \sigma_1, \sigma) \, d\sigma$ (the first order standard deviation) would be to use a weighted average of values of $\sigma$, say, for a simple case of one-order stochastic volatility:

$$\sigma (1 \pm a(1)), \ 0 \leq a(1) < 1$$

where *a(1) is* the proportional mean absolute deviation for $\sigma$, in other word the measure of the absolute error rate for $\sigma$. We use $\frac{1}{2}$ as the probability of each state.

Thus the distribution using the first order stochastic standard deviation can be expressed as:

$$f(x)_1 = \frac{1}{2} \{\phi(\mu, \sigma(1 + a(1)), x) + \phi(\mu, \sigma(1 - a(1)), x)\} \qquad (3)$$

***Illustration of the Convexity Effect:*** Figure 1 shows the convexity effect of a(1) for a probability of exceeding the deviation of x=6. With a[1]=$\frac{1}{5}$, we can see the effect of multiplying the probability by 7.

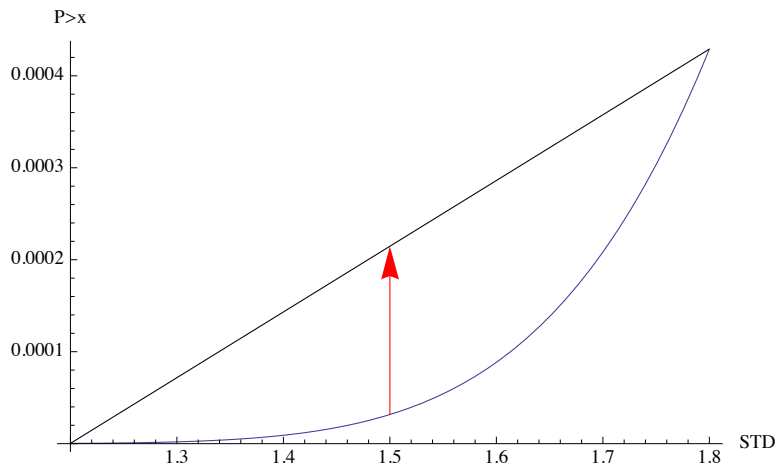

Figure 1 Illustration of he convexity bias for a Gaussian raising small probabilities: The plot shows the STD effect on P>x, and compares P>6 with a STD of 1.5 compared to P> 6 assuming a linear combination of 1.2 and 1.8 (here a(1)=1/5).

Now assume uncertainty about the error rate a(1), expressed by a(2), in the same manner as before. Thus in place of a(1) we have $\frac{1}{2}$ a(1)( 1± a(2)).

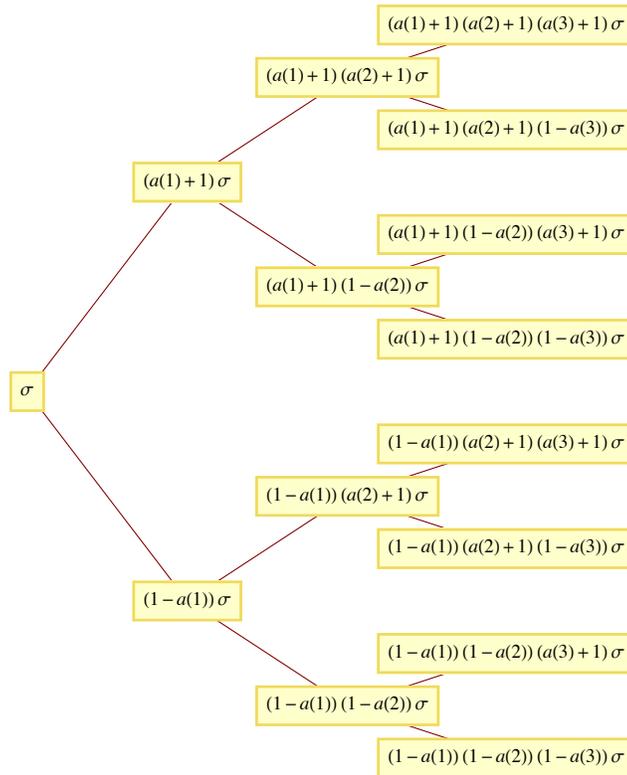

Figure 2- Three levels of error rates for $\sigma$ following a multiplicative process

The second order stochastic standard deviation:

$$f(x)_2 = \frac{1}{4} \{\phi(\mu, \sigma(1 + a(1)(1 + a(2))), x) + \phi(\mu, \sigma(1 - a(1)(1 + a(2))), x) + \phi(\mu, \sigma(1 + a(1)(1 - a(2))), x) + \phi(\mu, \sigma(1 - a(1)(1 - a(2))), x)\} \quad (4)$$

and the $N^{th}$ order:

$$f(x)_N = \frac{1}{2^N} \sum_{i=1}^{2^N} \phi(\mu, \sigma M_i^N, x) \quad (5)$$

where $M_i^N$ is the $i^{th}$ scalar (line) of the matrix $M^N$ $(2^N \times 1)$

$$M^N = \left\{ \prod_{j=1}^{N} (a(j) \, T[\![i, j]\!] + 1) \right\}_{i=1}^{2^N} \quad (6)$$

and T[[i,j]] the element of $i^{th}$ line and $j^{th}$ column of the matrix of the exhaustive combination of N-Tuples of (-1,1), that is the N-dimentional vector {1,1,1,...} representing all combinations of 1 and -1.
for N=3

$$T = \begin{pmatrix} 1 & 1 & 1 \\ 1 & 1 & -1 \\ 1 & -1 & 1 \\ 1 & -1 & -1 \\ -1 & 1 & 1 \\ -1 & 1 & -1 \\ -1 & -1 & 1 \\ -1 & -1 & -1 \end{pmatrix} \text{ and } M^3 = \begin{pmatrix} (1-a(1))(1-a(2))(1-a(3)) \\ (1-a(1))(1-a(2))(a(3)+1) \\ (1-a(1))(a(2)+1)(1-a(3)) \\ (1-a(1))(a(2)+1)(a(3)+1) \\ (a(1)+1)(1-a(2))(1-a(3)) \\ (a(1)+1)(1-a(2))(a(3)+1) \\ (a(1)+1)(a(2)+1)(1-a(3)) \\ (a(1)+1)(a(2)+1)(a(3)+1) \end{pmatrix}$$

so $M_1^3 = \{(1-a(1))(1-a(2))(1-a(3))\}$, etc.

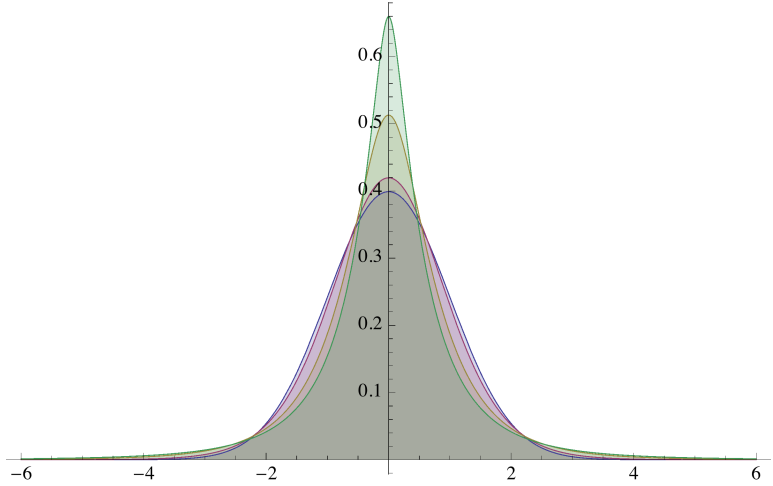

Figure 3, Thicker tails (higher peaks) for higher values of N; here N=0,5,10,25,50, all values of a=$\frac{1}{10}$

A remark seems necessary at this point: the various error rates *a(i)* are not similar to sampling errors, but rather projection of error rates into the future.

**Note**: we are assuming here, that $\sigma$ is stochastic with steps $(1 \pm a(n))$, not $\sigma^2$. An alternative method would be the mixture with a "low" variance $(\sigma(1-v))^2$ and a "high" one $\left(\sigma \sqrt{-v^2 + 2v + 1}\right)^2$ selecting a single *v* so that $\sigma^2$ remains the same in expectation. With $1 > v \geq 0$, the total standard deviation.

### *The Final Mixture Distribution*

The mixture weighted average distribution (recall that $\phi$ is the ordinary Gaussian with mean $\mu$, std $\sigma$ and the random variable *x*).

$$f(x \mid \mu, \sigma, M, N) = 2^{-N} \sum_{i=1}^{2^N} \phi\left(\mu, \sigma M_i^N, x\right) \tag{7}$$

**Note**: It could be approximated by a lognormal distribution for $\sigma$ and the corresponding V as its own variance. But it is precisely the V that interest us, and V depends on how higher order errors behave.

Next let us consider the different regimes for higher order errors.

## Regime 1 (Explosive): Case of a Constant parameter *a*

**Special case of constant *a***: Assume that a(1)=a(2)=...a(N)=a, i.e. the case of flat proportional error rate *a*. The Matrix *M* collapses into a conventional binomial tree for the dispersion at the level *N*.

$$f(x \mid \mu, \sigma, M, N) = 2^{-N} \sum_{j=0}^{N} \binom{N}{j} \phi\left(\mu, \sigma(a+1)^j(1-a)^{N-j}, x\right) \tag{8}$$

Because of the linearity of the sums, when a is constant, we can use the binomial distribution as weights for the moments (note again the artificial effect of constraining the first moment $\mu$ in the analysis to a set, certain, and known *a priori*).

$$\begin{array}{c}
\text{Moment} \\
\mu \\
\sigma^2 (a^2 + 1)^N + \mu^2 \\
3\mu\sigma^2(a^2+1)^N + \mu^3 \\
6\mu^2\sigma^2(a^2+1)^N + \mu^4 + 3(a^4+6a^2+1)^N \sigma^4 \\
10\mu^3\sigma^2(a^2+1)^N + \mu^5 + 15(a^4+6a^2+1)^N \mu\sigma^4 \\
15\mu^4\sigma^2(a^2+1)^N + \mu^6 + 15((a^2+1)(a^4+14a^2+1))^N \sigma^6 + 45(a^4+6a^2+1)^N \mu^2 \sigma^4 \\
21\mu^5\sigma^2(a^2+1)^N + \mu^7 + 105((a^2+1)(a^4+14a^2+1))^N \mu\sigma^6 + 105(a^4+6a^2+1)^N \mu^3 \sigma^4 \\
t^6 \sigma^2(a^2+1)^N + \mu^8 + 105(a^8+28a^6+70a^4+28a^2+1)^N \sigma^8 + 420((a^2+1)(a^4+14a^2+1))^N \mu^2\sigma^6 + 210(a^4+6a^2+
\end{array}$$

For clarity, we simplify the table of moments, with $\mu=0$

$$\begin{pmatrix}
\text{Order} & \text{Moment} \\
1 & 0 \\
2 & (a^2+1)^N \sigma^2 \\
3 & 0 \\
4 & 3(a^4+6a^2+1)^N \sigma^4 \\
5 & 0 \\
6 & 15(a^6+15a^4+15a^2+1)^N \sigma^6 \\
7 & 0 \\
8 & 105(a^8+28a^6+70a^4+28a^2+1)^N \sigma^8
\end{pmatrix}$$

Note again the oddity that in spite of the explosive nature of higher moments, the expectation of the absolute value of x is both independent of *a* and *N*, since the perturbations of $\sigma$ do not affect the first absolute moment $\int |x| f(x) \, dx = \sqrt{\frac{2}{\pi}} \sigma$ (that is, the initial assumed $\sigma$). The situation would be different under addition of *x*.

Every recursion multiplies the variance of the process by $(1+a^2)$. The process is similar to a stochastic volatility model, with the standard deviation (not the variance) following a lognormal distribution, the volatility of which grows with M, hence will reach infinite variance at the limit.

### *Consequences*

For a constant $a > 0$, and in the more general case with variable a where a(n) ≥ a(n-1), the moments explode.

A- Even the smallest value of $a > 0$, since $(1+a^2)^N$ is unbounded, leads to the second moment going to infinity (though not the first) when N→∞. So something as small as a .001% error rate will still lead to explosion of moments and invalidation of the use of the class of $\mathcal{L}^2$ distributions.

B- In these conditions, we need to use power laws for epistemic reasons, or, at least, distributions outside the $\mathcal{L}^2$ norm, regardless of observations of past data.

Note that we need an *a priori* reason (in the philosophical sense) to cutoff the N somewhere, hence bound the expansion of the second moment.

*Convergence to Properties Similar to Power Laws*

We can see on the example next Log-Log plot (Figure 1) how, at higher orders of stochastic volatility, with equally proportional stochastic coefficient, (where a(1)=a(2)=...=a(N)= $\frac{1}{10}$ ) how the density approaches that of a power law (just like the Lognormal distribution at higher variance), as shown in flatter density on the LogLog plot. The probabilities keep rising in the tails as we add layers of uncertainty until they seem to reach the boundary of the power law, while ironically the first moment remains invariant.

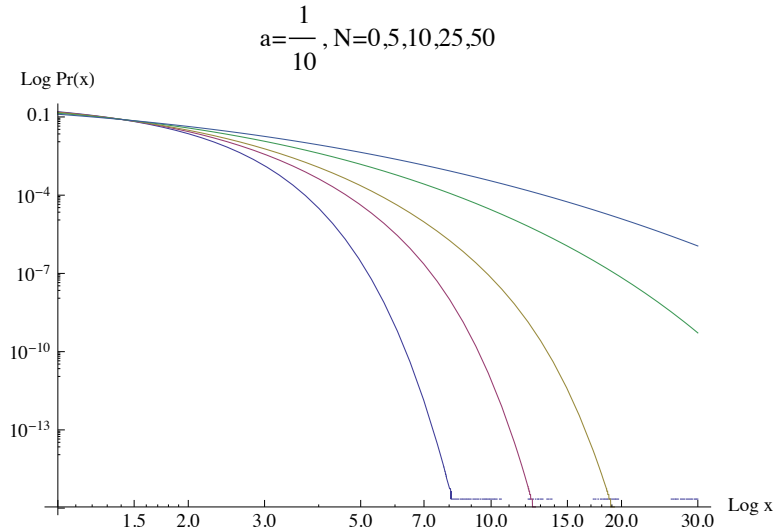

Figure x - LogLog Plot of the probability of exceeding x showing power law-style flattening as N rises. Here all values of a= 1/10

The same effect takes place as a increases towards 1, as at the limit the tail exponent P>x approaches 1 but remains >1.

*Effect on Small Probabilities*

Next we measure the effect on the thickness of the tails. The obvious effect is the rise of small probabilities.

Take the exceedant probability, that is, the probability of exceeding $K$, given $N$, for parameter $a$ constant :

$$P > K \mid N = \sum_{j=0}^{N} 2^{-N-1} \binom{N}{j} \text{erfc}\left(\frac{K}{\sqrt{2}\ \sigma\ (a+1)^{j}\ (1-a)^{N-j}}\right) \tag{9}$$

where erfc(.) is the complementary of the error function, 1-erf(.), erf(z) = $\frac{2}{\sqrt{\pi}} \int_0^z e^{-t^2}\, dt$

**Convexity effect**: The next Table shows the ratio of exceedant probability under different values of N divided by the probability in the case of a standard Gaussian.

$a = \frac{1}{100}$

| N | $\frac{P>3,N}{P>3,N=0}$ | $\frac{P>5,N}{P>5,N=0}$ | $\frac{P>10,N}{P>10,N=0}$ |
|---|---|---|---|
| 5 | 1.01724 | 1.155 | 7 |
| 10 | 1.0345 | 1.326 | 45 |
| 15 | 1.05178 | 1.514 | 221 |
| 20 | 1.06908 | 1.720 | 922 |
| 25 | 1.0864 | 1.943 | 3347 |

$a = \frac{1}{10}$

| N | $\frac{P>3,N}{P>3,N=0}$ | $\frac{P>5,N}{P>5,N=0}$ | $\frac{P>10,N}{P>10,N=0}$ |
|---|---|---|---|
| 5 | 2.74 | 146 | $1.09 \times 10^{12}$ |
| 10 | 4.43 | 805 | $8.99 \times 10^{15}$ |

| 15 | 5.98 | 1980 | $2.21 \times 10^{17}$ |
| 20 | 7.38 | 3529 | $1.20 \times 10^{18}$ |
| 25 | 8.64 | 5321 | $3.62 \times 10^{18}$ |

# Regime 2: Cases of decaying parameters *a*(*n*)

As we said, we may have (actually we need to have) *a priori* reasons to decrease the parameter *a* or stop *N* somewhere. When the higher order of *a*(i) decline, then the moments tend to be capped (the inherited tails will come from the lognormality of $\sigma$).

### *Regime 2-a; First Method: "bleed" of higher order error*

Take a "bleed" of higher order errors at the rate $\lambda$, $0 \leq \lambda < 1$, such as $a(N) = \lambda\ a(N-1)$, hence $a(N) = \lambda^N a(1)$, with $a(1)$ the conventional intensity of stochastic standard deviation. Assume $\mu=0$.

With *N*=2, the second moment becomes:

$$M2(2) = \left(a(1)^2 + 1\right)\sigma^2 \left(a(1)^2 \lambda^2 + 1\right) \tag{10}$$

With *N*=3,

$$M2(3) = \sigma^2 \left(1 + a(1)^2\right)\left(1 + \lambda^2 a(1)^2\right)\left(1 + \lambda^4 a(1)^2\right) \tag{11}$$

finally, for the general N:

$$M3(N) = \left(a(1)^2 + 1\right)\sigma^2 \prod_{i=1}^{N-1}\left(a(1)^2 \lambda^{2i} + 1\right) \tag{12}$$

We can reexpress (12) using the $Q$ – Pochhammer symbol $(a; q)_N = \prod_{i=1}^{N-1}\left(1 - aq^i\right)$

$$M2(N) = \sigma^2 \left(-a(1)^2; \lambda^2\right)_N \tag{13}$$

Which allows us to get to the limit

$$\text{Limit } M2\ (N)_{N \to \infty} = \sigma^2\ \frac{(\lambda^2; \lambda^2)_2\ (a(1)^2; \lambda^2)_\infty}{(\lambda^2 - 1)^2\ (\lambda^2 + 1)} \tag{14}$$

As to the fourth moment:

By recursion:

$$M4(N) = 3\ \sigma^4 \prod_{i=0}^{N-1}\left(6\ a(1)^2\ \lambda^{2i} + a(1)^4\ \lambda^{4i} + 1\right) \tag{15}$$

$$M4(N) = 3\ \sigma^4 \left(\left(2\sqrt{2} - 3\right)a(1)^2; \lambda^2\right)_N \left(-\left(3 + 2\sqrt{2}\right)a(1)^2; \lambda^2\right)_N \tag{16}$$

$$\text{Limit } M4(N)_{N \to \infty} = 3\ \sigma^4 \left(\left(2\sqrt{2} - 3\right)a(1)^2; \lambda^2\right)_\infty \left(-\left(3 + 2\sqrt{2}\right)a(1)^2; \lambda^2\right)_\infty \tag{17}$$

So the limiting second moment for $\lambda$=.9 and a(1)=.2 is just 1.28 $\sigma^2$, a significant but relatively benign convexity bias. The limiting fourth moment is just `9.88` $\sigma^4$, more than 3 times the Gaussian's (3 $\sigma^4$), but still finite fourth moment. For small values of a and values of $\lambda$ close to 1, the fourth moment collapses to that of a Gaussian.

### *Regime 2-b; Second Method, a Non Multiplicative Error Rate*

For N recursions

$$\sigma\ (1 \pm (a(1)(1 \pm (a(2)(1 \pm a(3)(\ ...))))$$

$$P(x, \mu, \sigma, N) = \frac{\sum_{i=1}^{L} f(x, \mu, \sigma (1 + (T^N . A^N)_i))}{L} \tag{18}$$

$(M^N . T + 1)_i$ is the $i^{th}$ component of the $(N \times 1)$ dot product of $T^N$ the matrix of Tuples in (6), $L$ the length of the matrix, and $A$ is the vector of parameters

$$A^N = \{a^j\}_{j=1,...N}$$

So for instance, for $N=3$, $T = \{1, a, a^2, a^3\}$

$$T^3 . A^3 = \begin{pmatrix} a + a^2 + a^3 \\ a + a^2 - a^3 \\ a - a^2 + a^3 \\ a - a^2 - a^3 \\ -a + a^2 + a^3 \\ -a + a^2 - a^3 \\ -a - a^2 + a^3 \\ -a - a^2 - a^3 \end{pmatrix}$$

The moments are as follows:

$$M1(N) = \mu \tag{19}$$

$$M2(N) = \mu^2 + 2\sigma \tag{20}$$

$$M4(N) = \mu^4 + 12 \mu^2 \sigma + 12 \sigma^2 \sum_{i=0}^{N} a^{2i} \tag{21}$$

at the limit of $N \to \infty$

$$\lim_{N \to \infty} M4(N) = \mu^4 + 12 \mu^2 \sigma + 12 \sigma^2 \frac{1}{1 - a^2} \tag{22}$$

which is very mild.

## Conclusions and Open Questions

So far we examined two regimes, one in which the higher order errors are proportionally constant, the other one in which we can allow them to decline. The difference between the two is easy to spot: the first category corresponds to naturally thin-tailed domains (higher errors decline rapidly), something very rare on mother earth. Outside of these very special situations (say in some strict applications or clear cut sampling problems from a homogeneous population, or similar matters stripped of higher order uncertainties), the Gaussian and its siblings (along with the measures such as STD, correlation, etc.) should be completely abandoned, along with any attempt to measure small probabilities. So the power law distributions are to be used more prevalently than initially thought.

- Can we separate the two domains along the rules of tangibility/subjectivity of the probabilistic measurement? Daniel Kahneman had a saying about measuring future states: how can one "measure" something that does not exist? So we could use:

  - Regime 1: elements entailing forecasting and "measuring" future risks. So should we use time as a dividing criterion: Anything that has time in it (meaning involves a forecast of future states) needs to fall into the first regime of non-declining proportional uncertainty parameters a(i).

  - Regime 2: conventional statistical measurements of matters patently thin-tailed, say as in conventional sampling theory, with a strong *a priori* acceptance of the methods without any form of skepticism.

- We can even work backwards, using the behavior of the estimation errors a(n) < a(1) or a(n) ≥ a(1) as a way to separate uncertainties.

*Note 1*

Infinite variance is not a problem at all -- yet economists have been historically scared of it. All we have to do is avoid using variance and measures in the $\mathcal{L}^2$ norm. For instance we can do much of what we currently do (even price financial derivatives) by using mean absolute deviation of the random variable, E[|x|] in place of $\sigma$, so long as the tail exponent of the power law exceeds 1 (Taleb, 2008).

## *Note 2*

There is most certainly a cognitive dimension, rarely (or, I believe, never) addressed or investigated, in the following mental shortcomings that, from the research, appears to be common among probability modelers:

- Inability (or, perhaps, as the cognitive science literature seems to now hold, lack of motivation) to perform higher order recursions among a certain class of people (*I know that he knows that I know that he knows*...). See the second edition of *The Black Swan*, Taleb (2010).

- Inability (or lack of motivation) to transfer from one situation to another (similar to the problem of weakness of central coherence). For instance, a researcher can accept power laws in one domain yet not recognize them in another, not integrating the ideas (lack of central coherence). I have observed this total lack of central coherence with someone who can do stochastic volatility models but is unable to understand them outside the exact same conditions when doing other papers.

## Acknowledgments

Jean-Philippe Bouchaud, Raphael Douady, Charles Tapiero, Aaron Brown, Dana Meyer, Andrew Gelman, Felix Salmon.